\documentclass[aps,prl,twocolumn,superscriptaddress,showpacs]{revtex4}
\usepackage{amsfonts,amscd,amsmath}
\newcommand{\oh}{\frac{1}{2}}
\def\ep{\text{e}}

\def\4{\tfrac{1}{4}}

\def\g{\mathfrak{g}}

\begin{document}
\preprint{LMU-ASC 13/07}
\title{Gluon Condensate, Wilson Loops and Gauge/String Duality}
\author{Oleg Andreev}
\affiliation{Max-Planck Institut f\" ur Physik, F\" ohringer Ring 6, 80805 M\" unchen, Germany}
\affiliation{L.D. Landau Institute for Theoretical Physics, Kosygina 2, 119334 Moscow, Russia}

\author{Valentin I. Zakharov}
\affiliation{Istituto Nazionale di Fisica Nucleare -- Sezione di Pisa, Largo Pontecorvo 3, 56127 Pisa, Italy}
\affiliation{Max-Planck Institut f\" ur Physik, F\" ohringer Ring 6, 80805 M\" unchen, Germany}
\begin{abstract}
We test  gauge/string duality by evaluating expectation values of small Wilson loops in pure Yang-Mills theories.  On the gauge-theory 
side, there exists rich phenomenology. The dual formulation provides with a universal language to evaluate the gluon condensate and 
quadratic correction in terms of the metric in the fifth coordinate. Quantitatively, the estimated value of the gluon condensate is 
approximately $0.010\,\text{GeV}^4$.
\end{abstract}
\maketitle

{\it Introduction.} The determination of the condensates of the QCD vacuum is a very important issue in phenomenology  of strong
interactions \cite{svz}. As known, the condensates can only be determined in a non-perturbative formulation of the theory. Until recently, the
lattice formulation was in this respect a unique theoretical tool. There is a long history of attempts to determine the gluon condensate
from first principles by computing
the expectation value of the gluonic action and subtracting the perturbative
contribution \cite{lattice,rakow}. One needs, however, to subtract many orders of perturbation theory, and the interpretation becomes
less straightforward. Thus, there is a strong need for new
alternative approaches to the problem.

This subject has taken an interesting turn with Maldacena duality \cite{malda}. Although the original proposal was for conformal theories,
various modifications have been found that produce gauge/string duals with a mass gap, confinement, and supersymmetry
breaking \cite{ads}.

In this paper we will address the issues of the gluon condensates
and of the quadratic correction  in a dual formulation, known as AdS/QCD or
holographic QCD.

{\it The model.} Let us first explain the model to be considered.  The dual string spacetime is the product of five-dimensional space, with
the Euclidean metric \cite{ans}

\begin{equation}\label{metric}
ds^2=\ep^{\oh cz^2}\frac{R^2}{z^2}\left(dx^idx^i+dz^2\right)
\,,
\end{equation}
and some five dimensional internal compact space $X$. We also take a constant dilaton. The AdS/QCD approach is a simplified version of
duality that assumes a trivial dependence on the internal space $X$ \cite{x}. Unlike other duals, this model does share a few key
features with QCD that singles it out and makes it very attractive for phenomenology: First, it is a nearly conformal theory at UV. Second,
it results in linear Regge-like spectra for mesons \cite{son, oa}. This fact allows one to fix the value of $c$. It is of order $0.9\,\text{GeV}^2$ 
\cite{comm}. Finally, the model results in a phenomenologically satisfactory description  of the confining potential as well \cite{az}. Note that 
the model does not contain any free fit parameter. Thus, evaluation of the gluon condensate we are going to undertake can be considered as a next
crucial test of the model.

Given the metric,  it is straightforward to calculate the expectation value of a Wilson loop by using the prescription of AdS/CFT \cite{ads}. We
first need to find the minimal area of string worldsheet which ends on a loop $\cal C$ living on the boundary ($z=0$) of spacetime. Then the
expectation value of the loop is given by

\begin{equation}\label{Wilson}
\langle\,W({\cal C})\,\rangle\simeq\ep^{-S}
\,.
\end{equation}
Finally,  the gluon condensate introduced in \cite{svz}

\begin{equation}\label{G2}
G_2=\Bigl\langle\,\frac{\alpha_s}{\pi}G_{\mu\nu}^aG_{\mu\nu}^a\,\Bigr\rangle
\,.
\end{equation}
is given by the coefficient of $s^2$, with $s$ the area of the loop, in the expansion of the
Wilson loop as $s\rightarrow 0$. Note that higher dimensional condensates are determined as coefficients of $s^n$ for $n>2$.

{\it Calculating the circular Wilson loop.} Now, let $\cal C$ be a circular loop of radius $a$. The worldsheet area is given by the Nambu-Goto
action $S=\frac{1}{2\pi\alpha'}\int d^2\xi\,\sqrt{\gamma}$, with $\gamma_{ab}$ the induced metric on the worldsheet. We choose
$\xi^1=r$ and $\xi^2=\varphi$, where $(r,\varphi )$ are the polar coordinates on a plane  in $R^4$. After introducing a new variable
$t=\sqrt{1-(\tfrac{r}{a})^2}$ and setting $\psi=(\tfrac{z}{a})^2$, the action becomes

\begin{equation}\label{ng2}
S=\g \int^{1}_0\,
dtt\,\psi^{-\tfrac{3}{2}}\,\ep^{\lambda\psi}\sqrt{\psi+\frac{1-t^2}{4t^2}(\psi')^2}
\,,
\end{equation}
where $\g=\tfrac{R^2}{\alpha'}$ and $\lambda=\frac{1}{2}ca^2$. A prime denotes a derivative with respect to $t$. The Euler-Lagrange
equation coming from this action is

\begin{equation}\label{eqm}
\begin{split}
&\psi\psi''-\frac{1}{t}\psi\psi'-\lambda\psi\Bigl((\psi')^2+\frac{4t^2}{1-t^2}\psi\Bigr)\\
&+
\Bigl(1-\frac{1}{2t}\psi'\Bigr)\Bigl(\frac{1}{2}(\psi')^2+\frac{4t^2}{1-t^2}\psi\Bigr)
=0
\,.
\end{split}
\end{equation}

We are interested in the physical situation where the loop may be considered as relatively small, with $a$ of order $0.05\,\text{ fm}$. This means
that the value of $\lambda$ is of order $0.03$. Thus, we can expand $\psi$ in powers $\lambda$ such that $\psi=\sum_{n=0}\psi_n\lambda^n$,
with the $\psi_n$'s obeying a set of differential equations.

In particular, $\psi_0$ is determined from

\begin{equation}\label{gross}
\psi_0\psi''_0-\frac{1}{t}\psi_0\psi'_0+
\Bigl(1-\frac{1}{2t}\psi'_0\Bigr)\Bigl(\frac{1}{2}(\psi'_0)^2+\frac{4t^2}{1-t^2}\psi_0\Bigr)=0
\,.
\end{equation}
This equation has the first integral

\begin{equation}\label{integral}
I=(1-t^2)(\psi_0)^{-\oh}\Bigl(\psi_0+\frac{1-t^2}{4t^2}(\psi'_0)^2\Bigr)^{-\oh}
\Bigl(1-\frac{1}{2t}\psi'_0\Bigr)
\,.
\end{equation}
We look for solutions of \eqref{gross} which are regular and non-vanishing at $t=1$. It follows from \eqref{integral} that $I=0$ on
this class of solutions. Since $\psi_0$ is positive for $t\leq 1$, the non-linear equation \eqref{integral} reduces to a simple linear differential
equation

\begin{equation}\label{lineq}
2t-\psi'_0=0
\,
\end{equation}
whose solution with the boundary condition $\psi_0(0)=0$ is simply

\begin{equation}\label{gross-sol}
\psi_0(t)=t^2
\,.
\end{equation}
This solution was found in \cite{gross, malda2}. In the original variables it takes the form $z=\sqrt{a^2-r^2}$ and describes the minimal
surface in $\text{AdS}_5$ bounded by a circle of radius $a$. Note that it can be obtained in a simpler way based on the underlying conformal
symmetry of the problem \cite{malda2}.

Now, we want to find corrections to $\psi_0$. This is nothing but a deformation of the minimal surface \eqref{gross-sol} due to the exponential
factor in the metric \eqref{metric}. For our purposes, it is sufficient to consider $\psi_1$ and $\psi_2$.

We begin with the leading correction. Using \eqref{eqm} and \eqref{gross-sol}, we find that $\psi_1$ obeys the following linear differential
equation

\begin{equation}\label{psi1}
\left(1-t^2\right)\psi''_1-\frac{2}{t}\psi'_1-4t^2=0
\,.
\end{equation}
Again, we look for a solution which is regular at $t=1$ and satisfies $\psi_1(0)=0$. This requirement is met by
fixing the two free parameters of a general solution (solution to $(1-t^2)\psi''_1-(2/t)\psi'_1=0$). Finally, $\psi_1$ takes
the form

\begin{equation}\label{psi1sol}
\psi_1(t)=4t-2t^2-4\ln(1+t)
\,.
\end{equation}

Now we move on to the next-to-leading correction. In this case equation \eqref{psi1} is replaced by

\begin{equation}\label{psi2}
\left(1-t^2\right)\biggl(\psi''_2-\Bigl(4t-\frac{1}{t^3}\psi_1+\frac{1}{t^2}\psi_1'\Bigl)\psi_1'\biggr)
-\frac{2}{t}\psi'_2
-4t^2\psi_1
=0,
\end{equation}
where $\psi_1$ is given by \eqref{psi1sol}.

As before, we look for a solution that is regular at $t=1$ and vanishing at $t=0$. A simple algebra shows that this requirement allows one to fix the two free parameters of a general solution. As a result, we have

\begin{multline}\label{psi2sol}
\psi_2(t)=16\,\text{Li}_2\Bigl(\frac{1+t}{2}\Bigr)+8\Bigl(\frac{1}{3}-4\ln2\Bigr)t-\frac{28}{3}t^2-\frac{2}{3}t^4\\
+8\ln(1+t)\biggl(\ln(1+t)+
2\ln(1-t)+t^2+t-\frac{1}{3}\biggr)\\-16\,\ln2\ln(1-t)-\frac{4}{3}\pi^2+8\ln^22
\,.
\end{multline}
Here $\text{Li}_2$ is the dilogarithm function. We have used that $\text{Li}_2(\tfrac{1}{2})=\tfrac{1}{12}\pi^2-\tfrac{1}{2}\ln^22$.

Having found the minimal surface, we can now calculate the corresponding area. Like $\psi$, it can also be expanded in powers $\lambda$ such
that $S=\sum_{n=0}S_n\lambda^n$.

To begin with, let us consider the leading contribution to $S$. It follows from \eqref{gross-sol} that it is simply given by

\begin{equation}\label{s0}
S_0(\epsilon)=\g\int^1_{\frac{\epsilon}{a}}\frac{dt}{t^2}=\g\Bigl(\frac{a}{\epsilon}-1\Bigr)
\,.
\end{equation}
As in \cite{gross, malda2}, there is a divergence in $S_0$. To make sense of the minimal area, we have regularized the integral. In doing so, we
cut the integral off at the lower bound. In terms of $r$, this means keeping only the part of the surface with
$r\leq a\sqrt{1-(\tfrac{\epsilon}{a})^2}$. After subtracting the divergent term, we find

\begin{equation}\label{s0.1}
S_0=-\g
\,.
\end{equation}

Now we would like to go beyond the leading approximation and consider the corrections. The computation is straightforward but somewhat
lengthy. One can evaluate $S_i$ by reducing the integrals to elementary ones and boundary terms. In evaluating the boundary terms, it is
useful to use the following asymptotics: $\psi_1\rightarrow-\tfrac{4}{3}t^3$ and
$\psi_2\rightarrow\tfrac{4}{3}(4\ln2-\tfrac{17}{3})t^3$ as $t\rightarrow 0$. At the end of the day, we get

\begin{equation}\label{s1.2}
S_1=\frac{5}{3}\g
\,,\qquad
S_2=\kappa\g
\,,\qquad
\kappa=\frac{7}{3}\Bigl(\frac{17}{6}-4\ln2\Bigr)
\,.
\end{equation}
Here we set $\epsilon=0$ because the integrals converge. Note that $\kappa\approx 0.14$.

Having derived the leading and sub-leading corrections, we can now write down the expectation value of the circular Wilson loop of radius $a$.
In the approximation we are using, the result is given by

\begin{equation}\label{w}
W\simeq\exp\biggl\{\g\biggl(1-\frac{5}{6} ca^2-\frac{1}{4}\kappa c^2a^4+O(a^6)\biggr)\biggr\}
\,.
\end{equation}

{\it Qualitative features.}
Evaluation of the gluon condensate in terms of small Wilson loops is common
on the lattice, see \cite{lattice,rakow}.  Namely, one starts with the smallest
Wilson line possible, that is  plaquette and represents it as follows:
\begin{equation}\label{G2.0}
\ln W=-\sum_n c_n\alpha_s^n -\frac{\pi^2}{36}ZG_2 s^2+O(s^3)
\,,
\end{equation}
where the sum means the perturbative contribution
and the gluon condensate introduced in \cite{svz} is given by the coefficient
in front of
$s^2\,(s=\pi a^2)$.  Moreover, we put the number of colors $N_{c}=3$ and
the factor $Z$ is defined in terms of  the perturbative beta function
as $Z^{-1}=\beta/\beta^{(1)}(g)$, with $\beta^{(1)}$ the one-loop beta function.

Compare representations (\ref{G2.0}) and (\ref{w}) for the expectation value of the Wilson loop in the dual
and direct formulations of the Yang-Mills theories. There are terms proportional to $s^{2}$ in the both cases
and this allows for a unique determination of the gluon condensate, see below.

There is a subtle point: a dimension two correction that scales like $s$. It is obviously present in \eqref{w} but somewhat disguised
in \eqref{G2.0}, where it does appear due to the UV renormalons \cite{akza}.
Phenomenologically the quadratic correction is known to be related to
the confinement effects \cite{akza}. Unlike the gluon condensate, however,
this correction is not related to a matrix element of dimension two operator.
The dual formulation reveals  the universality of the quadratic correction,
relating it to form of the metric in the fifth dimension. This is to be considered as
a phenomenological success of the gauge/string duality.

The dual formulation settles also another mystery  of the lattice
determination of the gluon condensate. Namely, usually the $\Lambda_{QCD}^{4}$
term, or condensate is related to an infrared renormalon, or
factorial divergence of the perturbative series in (\ref{G2.0}).
However, the first 16 terms of the perturbative expansion
evaluated in Ref. \cite{rakow} do not exhibit any factorial divergence
while allowing for a consistent determination of the gluon condensate.
In the dual formulation, the gluon condensate is associated with
small values of the fifth coordinate $z$, see above. Thus, the dual formulation
avoids the problem of the infrared renormalon and provides a qualitative
explanation for the properties of the perturbative expansion in \eqref{G2.0}.

{\it Determining the gluon condensate.}
Now, the gluon condensate can be read off from \eqref{w}, and so

\begin{equation}\label{G2.1}
G_2=\frac{9}{\pi^4}\kappa\g Z^{-1} c^2
\,.
\end{equation}

Making an estimate requires some numerics. First, as in \cite{oa}, the value of $c$ is fixed from the slope of the Regge trajectory of $\rho(n)$
mesons. This gives $c=0.91\pm 0.011\,\text{GeV}^2$ if the resonances with $n>3$ are used. At this point the error is estimated to be less
than $2\%$. Next, the overall constant $\g$ is adjusted to fit the slope of the linear term of the Cornell potential \cite{az}. This assumes that
$(e\g/4\pi)c=1/\mathfrak{a}^2$, where $1/\mathfrak{a}^2$ is the slope of the Cornell potential. As known, it is a fitting parameter defined from
the quarkonia spectra. Using the values of \cite{cornell}, it is estimated to be $1/\mathfrak{a}^2=0.186\pm0.021\,\text{GeV}^2$. Putting all
this together, we find $\g=0.94\pm 0.11$. Here, the error is now of order $11\%$. At first glance, $1-Z^{-1}$ can be made arbitrary small
by letting the scale $a$ go to $0$. This is however in conflict with the fit of $1/\mathfrak{a}^2$ done within the range of $0.2-1\,\text{fm}$. What
saves the day is the fact that the linear term in the quark potential is valid at least up to $0.05\,\text{fm}$ \cite{badalian}. Then, the
size of the leading correction (which is RS independent) $\tfrac{1}{2\pi}\tfrac{\beta_1}{\beta_0}\alpha_s$ is $0.08-0.14$ depending on the
number of quarks. In our estimate of the condensate we will assume that $Z^{-1}\simeq 1$ within $20\%$ accuracy.

Assembling all the factors, we have

\begin{equation}\label{G2.2}
G_2=0.010\pm 0.0023\,\,\text{GeV}^4
\,.
\end{equation}
This is our main result. It is surprisingly close to the original phenomenological estimate $0.012\,\text{GeV}^4$ of \cite{svz}, though
somewhat smaller than another phenomenological estimate $0.024\,\text{GeV}^4$ of \cite{narison}. For comparison, the lattice calculations give bigger values like $0.04\,\text{GeV}^4$ \cite{rakow}.

{\it Conclusions.} Using AdS/QCD correspondence, we found corrections of dimension two and four
to the leading zero-point-fluctuations contribution to small Wilson loops.
 The estimated value of the gluon condensate is remarkably close to the phenomenological estimates.
The dual formulation provides also means for  universal description of the quadratic correction which has never been found in terms of the direct formulation. Our results suggest that a dual formulation might be very good for describing physics at strong coupling.

\begin{acknowledgments}
O.A. would like to thank G. Duplancic and P. Weisz for useful discussions. The work of O.A. was supported in part by DFG and by Russian Basic
Research Foundation Grant 05-02-16486.
\end{acknowledgments}


\begin{thebibliography}{0}
\bibitem{ans}
\eqref{metric} is the ansatz. This time we do not know equations that provide such a solution. As known, the equations are the condition for 
conformal invariance of a two-dimensional $\sigma$ model. Since $\alpha'/R^2$ is of order $1$ \cite{az}, one needs 
the exact beta-functions of background fields (Weyl-anomaly coefficients) rather than their leading $\alpha'$ expansions. 
This makes the problem extremely difficult. 
Instead of solving it, we follow "the inverse scattering problem": first, we suggest a solution, then we look for its phenomenological relevance.
\bibitem{x}
In the AdS/CFT duality geometric symmetries of $X$ are related with scalar fields on the gauge theory side. Since we consider a pure gauge theory 
without additional scalars, it seems reasonable to assume a trivial dependence on $X$.
\bibitem{comm}
The sign of the exponent in \eqref{metric} is different from that of \cite{son,oa}. The point is that we use the Euclidean metric. 
The Minkowskian metric of \cite{son,oa} is obtained via analytic continuation $\vec x\rightarrow i\vec x$ together with $z\rightarrow iz$.
\end{thebibliography}
\end{document}